\documentstyle[12pt]{article}

\begin{document}

%%%%%%%%%%%%%%%%%%%%%%%%%%%%%%%%%%%%%%%%%%%%%%%%%%%%%%%%%%%%%%%%%%%

\title{{\Huge Gravitation Theory with Propagating Torsion}
\thanks{Talk given at XI International
 Conference Problems in Quantum Field Theory,
 Dubna, Russia,  July 13-17,, 1998.}
}

\author{
{\normalsize P.~P.~Fiziev}\thanks{ E-mail:\,\, fiziev@phys.uni-sofia.bg\,\,\,
and (or)\,\,\,
fiziev@thsun1.jinr.ru}
\\
{\footnotesize Department of Theoretical Physics, Faculty of Physics,
Sofia University,}\thanks{Permanent address.}\\
{\footnotesize Boulevard~5 James Bourchier, Sofia~1164, Bulgaria }\\
{\footnotesize and}\\
{\footnotesize   Bogoliubov Laboratory of Theoretical Physics,
 Joint Institute for Nuclear Research,}\\
{\footnotesize   141980 Dubna, Moscow Region, Russia,}\\
}

\maketitle

\begin{abstract}
We present a review of some recent models of gravitation theory with
propagating torsion based on the use of a torsion-dilaton field and
propose one more model of this type which promises to be more realistic.
A proper universal self-consistent minimal action principle yields the
properties of this model and predicts the interactions of torsion-dilaton
field with the real matter.
The new model may be compatible with the string models with dilaton field and
gives a novel interpretation of the dilaton as a part of the space-time torsion.
A relation with some recent models of dilatonic gravity is also possible.
\end{abstract}

%%%%%%%%%%%%%%%%%%%%%%%%%%%%%%%%%%%%%%%%%%%%%%%%%%%%%%%%%%%%%%%%%%%
%\draft

\sloppy
%\scrollmode
%%%%%%%%%%%%%%%%%%%
\renewcommand{\baselinestretch}{1.3} %
\newcommand{\sla}[1]{{\hspace{1pt}/\!\!\!\hspace{-.5pt}#1\,\,\,}\!\!}
\newcommand{\db}{\,\,{\bar {}\!\!d}\!\,\hspace{0.5pt}}
\newcommand{\lambdab}{\,\,{\bar {}\!\!\lambda}\!\,\hspace{0.5pt}}
\newcommand{\partb}{\,\,{\bar {}\!\!\!\partial}\!\,\hspace{0.5pt}}
\newcommand{\dsla}{\partb}
\newcommand{\eql}{e _{q \leftarrow x}}
\newcommand{\eqr}{e _{q \rightarrow x}}
\newcommand{\ite}{\int^{t}_{t_1}}
\newcommand{\itz}{\int^{t_2}_{t_1}}
\newcommand{\itd}{\int^{t_2}_{t}}
\newcommand{\lfrac}[2]{{#1}/{#2}}
\newcommand{\sfrac}[2]{{\small \,\,\hbox{${\frac {#1} {#2}}$}}}
\newcommand{\dV}{d^4V\!\!ol}
\newcommand{\ben}{\begin{eqnarray}}
\newcommand{\een}{\end{eqnarray}}
\newcommand{\la}{\label}

%%%%%%%%%%%%%%%%%%%

\section{Introduction}
The affine geometry with torsion was invented by E. Cartan in 1922
\cite{Cartan}.
He suggested, too the idea to enlarge the framework of general relativity
using this geometry which is more general then Riemannian one.
Today the affine geometry has got many applications in
different physical theories. Here is a list of part of them:

1. Einstein-Cartan (EC) theories (See for example
\cite{HehlHeydeKerlick} -- \cite{HeCrMiNe} and the references there in);

2. Gauge theories of gravity -- Einstein-Cartan-Sciama-Kible
(ECSK) theories (See for example \cite{HehlHeydeKerlick} -- \cite{HeCrMiNe}
and the references there in);

3. Affine-metric theories of gravity (See for example the review
article \cite{HeCrMiNe} and the huge amount of references there in);

4. Theory of supergravity (see for example \cite{SabGas},
\cite{West} and the references there in);

5. All kinds of modern superstring theories (See for example
\cite{SS}, \cite{GSW}, the recent review article \cite{Kir}
and the references therein).

6. Theory of "strong gravity" (see for example \cite{SabGas},
and the references there in);

7. Theory of the gravitational singularities \cite{Esposito1},
\cite{Esposito2};

8. Theory of the plastic deformations in solid states
(see for example \cite{Schouten} -- \cite{Kleinert2},
and the references there in);

9. Theory of space-time defects (See for example
\cite{GaLet} -- \cite{Anandan}, and the references there in);

10. Theory of the Kustaanheimo-Stiefel transformation in
celestial mechanics \cite{StSch} and the corresponding extension in
quantum mechanics, especially, the Duru-Kleinert transformation
in calculation of the Feynman path integral for Colomb potential
(see for example \cite{Kleinert2} and the references there in);

11. Recently proposed new formulation of the very general relativity
in terms of teleparallel spaces \cite{Pereira1} -- \cite{Pereira3};

12. Low energy limit of string theory (See for example \cite{SS}, \cite{GSW},
\cite{Hammond1} -- \cite{Hammond3});
\vskip .2truecm
\noindent and so on. There exist a huge amount of papers on these subjects
and one may find the corresponding references in the literature, cited above.
Unfortunately, no well established {\em physical} results which manifest
the usefulness of the torsion in the fundamental physics were found up to now.
Therefore a lot of people do not believe in torsion.
Nevertheless at present a not very big scientific community still continue to
propose new models of gravity with torsion and to hope to realize after all
the fundamental Cartan idea as a part of modern development of physical
theory.

The purpose of this article is to represent the recent investigations of the model
of gravity with propagating torsion proposed by
A. Saa in \cite{Saa1} -- \cite{Saa5} which ware performed
by the author and his collaborators,
as far as some new models \cite{F1} -- \cite{FY2}
which are aimed to overcome the theoretical
and experimental inconsistencies of Saa's model.
Some of these new models are based on the previous researches on the action
principle in spaces with torsion \cite{Timan} -- \cite{Kl4}.
Finally we formulate one more new model of gravitation with propagating
torsion-dilaton which promises to be more realistic.

\section{The Self Consistent Minimal Coupling Principle Problem
         in Spaces with Torsion}

We shell start remaining some well known basic notions and definition from
differential geometry to adjust the terminology and notations we will use
further.
We denote as
${\cal M}^{(1,3)}\{g_{\alpha\beta}(x), \Gamma^\gamma_{\alpha\beta}(x)\}$\,\,
the affine connected space with connection coefficients
$\Gamma^\gamma_{\alpha\beta}(x)$
and metric tensor $g_{\alpha\beta}$ with signature (+,-,-,-).
This four-dimensional affine-metric space will be our model of
the physical space-time in what follows.
The connection and the metric in it further are supposed to be
related in general {\em only} by the metricity condition:
$\nabla_\alpha g_{\beta \gamma}\equiv 0$\footnote{Some times such affine-metric
spaces are called Einstein-Cartan manifolds
\cite{HehlHeydeKerlick} -- \cite{HeCrMiNe}.},
$\nabla_\alpha$ being the covariant derivatives with respect to the affine
connection with coefficients $\Gamma^\gamma_{\alpha\beta}$.
This metricity condition yields the important relation
$\Gamma^\gamma_{\alpha\beta}={\gamma \brace \alpha \beta}
+ {K_{\alpha\beta} }^\gamma$ where
${\gamma \brace \alpha \beta}= {\sfrac 1 2}g^{\gamma \mu}
(\partial_\alpha g_{\mu\beta} +\partial_\beta g_{\mu\alpha} -
\partial_\mu g_{\alpha\beta})$ are the Christoffel symbols,
i.e. the coefficients of the Levi-Civita connection,
${K_{\alpha\beta} }^\gamma =  {S_{\alpha\beta} }^\gamma
+ {S^\gamma}_{\alpha\beta} + {S^\gamma}_{\beta\alpha}$  is the contorsion
tensor, and ${S_{\alpha\beta} }^\gamma = \Gamma^\gamma_{[\alpha\beta]}$
is the torsion tensor (in coordinate basis). We shall need
the torsion vector $S_{\alpha} = {\sfrac 2 {D-1}}{S_{\alpha \mu}}^\mu =
{\frac 2 3}{S_{\alpha \mu}}^\mu$ for dimension $D=1+3$ of the space-time
\footnote{We use the Schouten's normalization conventions \cite{Schouten}.}.
As we see, due to the metricity condition the torsion enters into
the symmetric part of the connection coefficients, too:
$\Gamma^\gamma_{\{\alpha\beta\}}={\gamma \brace \alpha \beta}+
{S^\gamma}_{\alpha\beta} + {S^\gamma}_{\beta\alpha}$, nevertheless
it was defined in a coordinate basis as an anti-symmetric part of these
coefficients.
As a result if the torsion tensor $S_{\alpha\beta\gamma}$ is not complete
anti-symmetric in its three indexes the following specific problem
(which we shell call ''the G-A problem'')
appears in the affine-metric spaces:

Consider first the free motion of a
relativistic particle with mass $m$ in a flat Minkowski space
${\cal E}^{(1,3)}\{\eta = (+1,-1,-1,-1), \Gamma=0\}$,
$\eta$ being its metric.
The action functional
\begin{equation}
{\cal A}[x]= -mc \int \sqrt{ \eta_{ij} dx^i dx^j} = -mc \int ds
\label{pA1}
\end{equation}
under standard action principle
\begin{equation}
\delta {\cal A} [x] = 0
\label{AP1}
\end{equation}
leads to the dynamical equations:
\begin{equation}
mc^2 {\frac {d^2x^i} {ds^2}}=0
\label{SL}
\end{equation}
solved by a straight lines with uniform velocity.
These {\em straightest} lines are at the same time {\em the shortest} ones
which connect the initial and the final position of the particle.

The {\em standard} minimal coupling principle (MCP) maps dynamical
equations (\ref{SL}) onto the equations of motion:
\ben
mc^2\left( {\frac {d^2 x^\gamma} {ds^2}} +  \Gamma_{\alpha \beta}^\gamma
{\frac {d x^\alpha} {ds}}{\frac {d x^\beta} {ds}} \right)=
mc^2 {\frac D {ds}}{\frac {d x^\gamma} {ds}}  = 0,
\la{AL}
\een
which describe the straightest lines, i.e. {\em the autoparallel lines}
(A-lines) in the space
${\cal M}^{(1,3)}\{g_{\alpha\beta}(x), \Gamma^\gamma_{\alpha\beta}(x)\}$,\,
${\frac D {ds}}$ being the absolute derivative with respect to the
affine connection,
and the action (\ref{pA1}) onto the action
\ben
{\cal A}[x]= -mc\int \sqrt{g_{\mu \nu}(x)) dx^\mu dx^\nu}= -mc\int ds.
\la{pA2}
\een
But under {\em standard} action principle (\ref{AP1}) this action functional
yields the geodesic line (G-line) equations of motion
which describe the lines with stationary length
(sometimes called "shortest lines", nevertheless in general this may be not
true) in the space
${\cal M}^{(1,3)}\{g_{\alpha\beta}(x), \Gamma^\gamma_{\alpha\beta}(x)\}$:
\ben
mc^2\left( {\frac {d^2 x^\gamma} {ds^2}} + { \gamma \brace \alpha \beta}
{\frac {d x^\alpha} {ds}}{\frac {d x^\beta} {ds}} \right) =
mc^2 {\frac D {ds}}{\frac {d x^\gamma} {ds}}
- 2 mc^2 {S^\gamma}_{\alpha \beta}
{\frac {d x^\alpha} {ds}}{\frac {d x^\beta} {ds}} = 0.
\la{GL}
\een
Obviously the autoparallel equation (\ref{AL}) means
a free motion of the test spinless particle in the space
${\cal M}^{(1,3)}\{g_{\alpha\beta}(x), \Gamma^\gamma_{\alpha\beta}(x)\}$
with zero absolute acceleration:
$a^\gamma = c^2 {\frac D {ds}}{\frac {d x^\gamma} {ds}} = 0$.
This is the most natural translation of the usual dynamics of
test free particle and corresponds to the very physical notion of
a "free test particle".

In contrast, the geodesic equations (\ref{GL}) imply in general\footnote{In
Riemannian spaces, as far as in affine connected spaces with
complete anti-symmetric in its three indexes torsion tensor
$S_{\alpha\beta\gamma}$ G-lines coincide with A-lines and
no G-A problem exists.} the unnatural law
of free motion:\,\, $m a^\gamma = {\cal F^\gamma}$.
Hence, we actually have to introduce a specific "torsion force"
${\cal F^\gamma} = 2 mc^2 {S^\gamma}_{\alpha \beta} u^\alpha u^\beta$
($u^\alpha = {\frac {d x^\alpha} {ds}}$ being the particle's four-velocity)
to compensate the natural torsion dependence of the dynamics in the space
${\cal M}^{(1,3)}\{g_{\alpha\beta}(x), \Gamma^\gamma_{\alpha\beta}(x)\}$
and to allow the free test particle to follow
the usual extreme of the classical action (\ref{pA2}).

The same problem we observe in field dynamics of classical fields with
different spin. In the simplest case of a massive spinless scalar field
with mass $m$ the flat Minkowski equation of motion
\ben
\eta^{\mu\nu}\partial_\mu \partial_\nu \phi + m^2 \phi = 0
\la{sfEF}
\een
being derivable via standard action principle from the action integral
\ben
{\cal A}[\phi(x)]= \int d^4x {\sfrac 1 2}\!
\left( \eta^{\mu \nu} \partial_\mu \phi \partial_\nu \phi - m^2 \phi^2 \right)
\la{sfAF}
\een
is mapped onto the {\em autoparallel-type} (A-type) equation
\ben
\Box \phi + m^2 \phi = 0
\la{sfEA}
\een
under the standard MCP which produces the action
\ben
{\cal A}[\phi(x)]= \int d^4x \sqrt{|g(x)|}  {\sfrac 1 2}\!
\left( g^{\mu \nu} \nabla_\mu \phi \nabla_\nu \phi - m^2 \phi^2 \right)
\la{sfA}
\een
in the space
${\cal M}^{(1,3)}\{g_{\alpha\beta}(x), \Gamma^\gamma_{\alpha\beta}(x)\}$
(for scalar field $\nabla_\mu \phi \equiv \partial_\mu \phi$).
Then the {\em standard} action principle yields the {\em geodesic-type} (G-type)
equation of motion
\ben
{\stackrel{{}^{\{\}}}{\Box}} \phi + m^2 \phi =
\Box \phi + m^2 \phi + 3 S^\mu \nabla_\mu \phi = 0.
\la{sfEG}
\een
Here we use the laplasian $\Box = g^{\alpha\beta} \nabla_\alpha \nabla_\beta$
and Laplas-Beltrami operator
${\stackrel{{}^{\{\}}}{\Box}}=
g^{\alpha\beta} \stackrel{{\{\}}}{\nabla}_\alpha \stackrel{{\{\}}}{\nabla}_\beta
={1\over \sqrt{|g|}}\partial_\mu\left(\sqrt{|g|}g^{\mu\nu}\partial_\nu\right)=
\Box + 3 S^\mu \nabla_\mu$
in the space
${\cal M}^{(1,3)}\{g_{\alpha\beta}(x), \Gamma^\gamma_{\alpha\beta}(x)\}$,
$\stackrel{{\{\}}}{\nabla}_\alpha$ being the covariant derivative with
respect to the Levi-Civita connection with coefficients
${\gamma \brace \alpha \beta}$\footnote{We shall use the mark $\{\}$
above the symbols to denote all objects: operators, quantities, e.t.c.
which correspond to the Levi-Civita connection in the space
${\cal M}^{(1,3)}\{g_{\alpha\beta}(x), \Gamma^\gamma_{\alpha\beta}(x)\}$.}.

The A-type equation (\ref{sfEA}) is essentially different from the G-type
equation (\ref{sfEG}) if the torsion vector $S_\alpha$ does not vanish
and this leads to the G-A problem in the case under consideration.
If we consider the affine connection as a fundamental object which
defines the very geometry of the space
${\cal M}^{(1,3)}\{g_{\alpha\beta}(x), \Gamma^\gamma_{\alpha\beta}(x)\}$
all equations of motion have to be written in terms of its absolute
derivatives. Then the third term $3 S^\mu \nabla_\mu \phi$
in the corresponding form of the equation (\ref{sfEG}) has to be considered
as a density of an additional force ${\cal F}= -3 S^\mu \nabla_\mu \phi$
caused by torsion.
It has to be introduced to compensate the natural torsion dependence of
the scalar field dynamics generated by the direct application of the MCP
to the special relativistic equation of motion (\ref{sfEF}) of spinless field.

The above paradox in the description of the free motion of test particles
and fields in affine connected spaces
${\cal M}^{(1,3)}\{g_{\alpha\beta}(x), \Gamma^\gamma_{\alpha\beta}(x)\}$
with nonzero torsion forces one to answer the following two basic questions:

1) What is more fundamental:

\hskip .5truecm $\bullet$ \,\,the free motion as a motion without external
forces of any nature, and hence, with zero absolute acceleration,
according to Newton law of inertia; or

\hskip .5truecm $\bullet$ \,\,the free motion as a motion governed by
geodesic type of equations of motion in accordance with the {\em standard}
action principle.

2) Do the {\em self consistent} minimal coupling principle (SCMCP) exist
which will yield the same results when applied in action principle and directly
in the equation of motion in flat Minkowski space.

Concerning the answer of the first question it is quite obvious that the
Newton law of inertia has a more profound {\em physical} character.

In addition it is interesting to note that as early as in \cite{Weyl}
a deep analysis of the origin of inertial forces,
their relations with gravitation and the possibility of
a geometric description of these two types of forces which look quite different
at first glance brings Weyl to the conclusion that the inertia and
the gravitation are to be determined by some {\em affine connection}.
In Weyl's analysis the free motion is just an autoparallel displacement of
the particle velocity with respect to this affine connection\footnote{In the
present article we use a different terminology reserving the term "geodesics"
for the lines with stationary length.}. But after all
without any physical motivation Weyl superimposes the "usual" condition
${\Gamma_{\alpha\beta} }^\gamma = {\Gamma_{\beta\alpha} }^\gamma$
and thus arrives to the standard Levi-Civita connection in general relativity.

The well known argument to chose the second alternative answering the first
question is the fact, that the action principle follows from quantum mechanics
as a fundamental principle for classical motion \cite{Dirac}, \cite{Feynman}.
But there is no guarantee that the quantum mechanics leads to the usual form
of action principle in affine connected spaces with nonzero torsion.
Moreover it is found that Feynman path integral leads to the
Schr\"odinger equation of autoparallel type in such spaces
\cite{Kleinert2}, \cite{F5}.

So, answering the first question it will be very hard to drop out
from the physical theory one of the two well established principles
which conflict in the presence of torsion.
The best thing we can do will be to give a positive answer
to the second question and this way to overcome the G-A problem.
There exist two possibilities to do this:
the first is to change the very variational principle,
the second one is to change properly the action functional
in spaces with torsion,
i.e. to modify the minimal coupling principle in presence of torsion.

The autoparallel motion of test particle in affine connected spaces was
proposed in \cite{Ponamarev} and derived from formally
modified variational principle as early as in \cite{Timan}.
It was based on the following {\em new postulate}:
\ben
\biggl( \delta_x {\frac d {dt}} - {\frac d {dt}} \delta_x \biggr) x^\alpha=
2 {S_{\mu\nu}}^\alpha \dot x^\mu \delta x^\nu
\la{ComRel_T}
\een
for the commutation relation between variations of paths $\delta_x$
and time-derivative ${\frac d {dt}}$ in spaces with torsion.

One has to add that in Weitzenb\"ock affine flat spaces with torsion
a new variational principle for classical particle trajectories was
{\em derived} recently \cite{F2}, \cite{F3}, \cite{F4}.
It leads after all to autoparallel motion of the particles and gives
a proper development of Kleinert's "quantum equivalence principle"
\cite{Kleinert2}, \cite{Kleinert5}, \cite{Kleinert6}.
Once more a formal modification of the variational principle in spaces with
torsion based on the relation (\ref{ComRel_T}) was reinvented independently
in \cite{Kl3}.
Very recently the autoparallel motion of nonrelativistic particle was derived
from proper generalization of Gauss' principle of least constraint
in \cite{Kl4}.

Nevertheless the success of the Timan's variational principle in reaching a
SCMCP for classical particles and fluids we have to note that it yields
some problems \cite{F3}, \cite{F4}, \cite{Nuno} which still are to be overcome.

The second possibility: the use of the usual variational principle
after a proper modification of the action functional
was successfully examined by A. Saa \cite{Saa1} -- \cite{Saa5} for classical
matter fields of of any kind: scalar field $\phi$, spinor field $\psi$,
electromagnetic field $A_\mu$, Yang-Mills fields ${\bf A}_\mu$, e.t.c.
but only in the special case when the torsion vector is potential:
\ben
S_{\alpha} = \nabla_\alpha \Theta \equiv \partial_\alpha \Theta.
\la{GrCond}
\een
We shall call the potential $\Theta$ of the torsion vector
{\em a torsion-dilaton field}.

Saa's model for relativistic fluids and particles and some of
its modifications are considered in the articles \cite{F1} -- \cite{FY2}.
Analogous to Saa's solution of the G-A problem for classical relativistic
particles was found in \cite{Kl7}.

So, at present we have no complete solution of the G-A problem and
one can't exclude the possibility for geodesic motion.
Therefore we have to take into account this type of motion, too.
The reasonable goal is to develop both conceptual possibilities
to the form which will admit a comparison with the experimental evidences,
or will recover their theoretical (in)consistency.

For example, there exist the following consistency problem in the affine
connected space
${\cal M}^{(1,3)}\{g_{\alpha\beta}(x), \Gamma^\gamma_{\alpha\beta}(x)\}$.
In the Riemannian space the geodesic equation (\ref{GL}) for test particles
with mass $m$ follows from the scalar field
equation (\ref{sfEG}) with the same mass in a semiclassical limit,
See for example \cite{Schr}.
One expects to see the same property in the case of nonzero
torsion in the space
${\cal M}^{(1,3)}\{g_{\alpha\beta}(x), \Gamma^\gamma_{\alpha\beta}(x)\}$, too.
But the naive generalization of the corresponding procedure does not lead to
the expected result.
Indeed, representing the field $\phi$ in a form $\phi= A \exp(i\varphi)$
with some real amplitude $A$ and real phase $\varphi$ we can write down
$\Box \phi = \phi\left( {\Box A \over A} -
g^{\alpha\beta} \partial_\alpha \varphi \partial_\beta \varphi \right)+
i{\phi\over A^2}
\nabla_\alpha \left(A^2 g^{\alpha\beta} \nabla_\beta \varphi \right)$.
Now the autoparallel equation (\ref{sfEA}) in the semiclassical limit
${\Box A \over A}\approx 0$ yields the eikonal equation
\ben
g^{\alpha\beta} \partial_\alpha \varphi \partial_\beta \varphi = m^2
\la{EIC}
\een
which seems to correspond to the Hamilton-Jacobi equation for classical
action function $S=\hbar \varphi$ of the {\em geodesic} equation (\ref{GL}),
not of the autoparallel one (\ref{AL}).
In addition we reach the autoparallel type of conservation law
$\nabla_\alpha j^\alpha=0$
for the current $j_\alpha = A^2 \nabla_\alpha\varphi$.

It turns out that the solution of this consistency problem dictates a definite
{\em new type of interaction} of the torsion-dilaton field $\Theta$ with
the mass terms like ${\frac1 2}m^2 \phi^2$, $m \bar\psi \psi$,...
in the corresponding field equations of the model \cite{F1}.
This interaction has a form ${\frac1 2}m^2 \phi^2\, e^{p_\phi \Theta}$,
$m \bar\psi \psi\,e^{p_\psi \Theta}$,...; $p_\phi$, $p_\psi$,...
being proper model-dependent integers (See for details \cite{F1}).
It ensures that the semiclassical limit of the A-type wave equations
for fields yields an A-type equations of motion for the corresponding
classical particles. Actually this is a new modification of the MCP needed
to make it coherent with semiclassical physics and it gives a definite
consequences which allow an experimental check.

\section{The Theoretical Inconsistency of the Strict Saa's Model
         for Matter Fields and Particles}

The main idea of Saa's model of gravity with propagating torsion is to replace
the usual volume element $d^4V\!ol= \sqrt{|g|} d^4 x$ in Einstein-Cartan
manifold with a new one: $d^4V\!ol_{Saa}= e^{-3\Theta}\sqrt{|g|} d^4 x$.
It was pointed out in \cite{F1} that the Saa's volume is covariantly
constant with respect to the transposed connection in the space
${\cal M}^{(1,3)}\{g_{\alpha\beta}(x), \Gamma^\gamma_{\alpha\beta}(x)\}$
with coefficients
$(\Gamma^T)_{\alpha\beta}^\gamma = \Gamma_{\beta\alpha}^\gamma$,
not with respect to the usual connection with coefficients
$\Gamma_{\alpha\beta}^\gamma$.
Therefore the Einstein-Cartan manifold with such volume was called
transposed-equi-affine and the  corresponding theory of gravity  --
transposed-equi-affine theory of gravity (TEATG).
Then if we put the new volume element in the action integrals for all matter
fields: $\phi, \psi, A_\mu, {\bf A}_\mu, ... $ with {\em standard} lagrangian
${\cal L}_{\phi, \psi, A_\mu, {\bf A}_\mu, ...}$:
\ben
{\cal A}[\phi, \psi, A_\mu, {\bf A}_\mu, ...] =
{\frac 1 c} \int {\cal L}_{\phi, \psi, A_\mu, {\bf A}_\mu, ...}
e^{-3\Theta}\sqrt{|g|} d^4 x
\la{AMF}
\een
we will have an A-type equations of motion for all these fields
\cite{Saa1}-\cite{Saa5}. Hence, Saa's modification of the volume element
leads to a SCMCP {\em for all matter fields}.

In the strict Saa's model $d^4V\!ol_{Saa}$ is the universal
volume element in the space
${\cal M}^{(1,3)}\{g_{\alpha\beta}(x), \Gamma^\gamma_{\alpha\beta}(x)\}$ and
we have to use it in all volume integrals. The use of this volume in the theory
of relativistic fluids unfortunately is not complete successful \cite{F1}.
Using the corresponding generalization of Gauss' law for fluid we reach
an A-type continuity equation:
\ben
\nabla_\alpha \left( \mu(x) u^\alpha(x) \right) = 0,
\la{ACE}
\een
$\mu(x)$ being the fluid mass density, $u^\alpha(x)$ being the components of the
fluid's four velocity. But the standard action principle for the fluid action
with Saa's volume:
\ben
{\cal A}_m =-{\frac 1 c}\int \left({\mu c^2 + \mu \Pi}\right)
e^{-3\Theta}\sqrt{|g|} d^4 x
\la{AFl}
\een
where $\Pi$ is the elastic potential energy of the fluid, yields
the following G-type equations of motion for relativistic fluid:
\ben
(\varepsilon + p) u^\beta {\stackrel{{\{\}}}{\nabla}}_\beta u_\alpha =
\left(\delta^\beta_\alpha - u_\alpha u^\beta \right)
{\stackrel{{\{\}}}{\nabla}}_\beta p
\la{GFluEM}
\een
which may be rewritten in a form
\ben
(\varepsilon + p) u^\beta \nabla_\beta u_\alpha =
\left(\delta^\beta_\alpha - u_\alpha u^\beta \right)
\biggl(\nabla_\beta p +
(\varepsilon + p)\nabla_\beta\Theta\biggr).
\la{GFluEM2}
\een
Here a torsion-force density
${\cal F}_\alpha =(\varepsilon + p)\left(\delta^\beta_\alpha - u_\alpha u^\beta \right)
\partial_\beta\Theta$ appears. Hence, for relativistic
fluid and particles Saa's idea to reach SCMCP using the new volume element
$d^4V\!ol_{Saa}$ fails. This forces us to re-evaluate the good and the bad
features of Saa's model and to look for its further modifications.

In addition Saa's model leads to a definite action for geometric
fields $g_{\alpha \beta}$ and ${S_{\alpha\beta}}^\gamma$.
In the spirit of its general idea we have to put in the action
integral the new volume element and to use Hilbert-Einstein-like lagrangian
one uses in Einstein-Cartan theories of gravity with torsion
\cite{HehlHeydeKerlick} -- \cite{HeCrMiNe}:
\ben
{\cal A}[g_{\alpha \beta}, {S_{\alpha\beta}}^\gamma] =
-{\frac c {2\kappa} } \int\, R\, e^{-3\Theta}\sqrt{|g|} d^4 x,
\la{AGF}
\een
where $R$ is the Cartan scalar curvature, i.e. the scalar curvature with
respect to the whole affine connection and $\kappa$ is the Einstein constant.
This way we do not need to introduce some new interaction constants related
with the torsion and the whole theory is determined by the usual properties
of the matter, i.e. no new ''charges'' appear, nevertheless we have new torsion
field degrees of freedom. Moreover, the action (\ref{AGF}) incorporates general
relativity and gives definite action for the torsion field.

In the special case when only spinless matter presents the affine connection
is semi-symmetric \cite{Schouten} with gradient torsion vector:
\ben
{S_{\alpha\beta}}^\gamma = S_{[\alpha}\delta_{\beta]}^\gamma
=\partial_{[\alpha}\Theta\,\delta_{\beta]}^\gamma.
\la{SST}
\een
Then the Cartan scalar curvature is
$$R=\stackrel{\{\}}{R} + 6\nabla_\mu S^\mu + 12 S_\mu S^\mu =
\stackrel{\{\}}{R} + 6\left({\stackrel{{}^{\{\}}}{\Box}} \Theta -
g^{\mu \nu} \partial_\mu \Theta \partial_\nu \Theta\right),$$
$\stackrel{\{\}}{R}$ being the Riemann scalar curvature
(of the Levi-Civita connection) and the usual variation of the action
(\ref{AGF}) yields the equations for geometrical fields \cite{Saa5}, \cite{F1}:
\ben
G_{\mu\nu} +\nabla_\mu \nabla_\nu \Theta - g_{\mu\nu}\Box\Theta =
{\sfrac \kappa {c^2}}T_{\mu\nu},\nonumber\\
\Box\Theta  = {\sfrac \kappa {c^2}}
\left({\cal L}_M -{\sfrac 1 3}{\delta{\cal L}_M \over \delta \Theta}\right)
-{\sfrac 1 2} R.
\la{GFE}
\een
where ${\cal L}_M$ is the lagrangian of the matter and matter fields.

\section{The Model with Variable Planck ``Constant''}

The simplest modification of Saa's model which preserves the SCMCP for matter
fields may be reach if we will use the volume
$d^4V\!ol_{Saa}= e^{-3\Theta}\sqrt{|g|} d^4 x$
{\em only in the action integrals}
and the usual volume $d^4V\!ol=\sqrt{|g|} d^4 x$ in all other formulae with
volume integrals.
Then we will have the following total action for geometric fields,
matter fields, fluids and particles:
\ben
{\cal A}_{total}[g_{\alpha \beta}, {S_{\alpha\beta}}^\gamma;
\phi, \psi, A_\mu, {\bf A}_\mu,...; \mu, m,...] = \nonumber\\
-{\frac c {2\kappa} } \int\,e^{-3\Theta}\, R \sqrt{|g|} d^4 x
+{\frac 1 c}\int\,e^{-3\Theta}\,{\cal L}_{\phi, \psi, A_\mu, {\bf A}_\mu,...}
\sqrt{|g|} d^4 x \nonumber \\
-{\frac 1 c}\int\,e^{-3\Theta}\, \left({\mu c^2 + \mu \Pi}\right)
\sqrt{|g|} d^4 x
-mc\int\,e^{-3\Theta}\,\sqrt{g_{\mu \nu}(x)) dx^\mu dx^\nu}.
\la{A_total}
\een
The presence of the factor $e^{-3\Theta}$ in all action
integrals simply means that instead of the usual Planck constant
we are introducing a Planck field
\ben
\hbar (x) = \hbar_\infty  e^{3\Theta(x)},
\la{Planck}
\een
$\hbar_\infty$ being the Planck constant in vacuum far from matter.
Indeed, we actually need the classical action functionals
just to calculate quantum transition amplitudes via the Feynman path
integral:
\ben
\int{\cal D}\left(
\vbox to 12pt{}
g_{\alpha \beta}, {S_{\alpha\beta}}^\gamma;
\phi, \psi, A_\mu, {\bf A}_\mu,...; \mu, m,...
\right) \times \nonumber \\
\exp\left({\frac i {\hbar_\infty}}
{\cal A}_{total}[g_{\alpha \beta}, {S_{\alpha\beta}}^\gamma;
\phi, \psi, A_\mu, {\bf A}_\mu,...; \mu, m,...]
\right).
\la{QA}
\een
Now it is obvious that the very Planck constant $\hbar_\infty$ may be
included in the factor $e^{-3\Theta(x)}$, but more important is the
observation that we must do this, because the presence of this {\em uniform}
factor in the formula (\ref{QA}) means that we actually introduce a local
Planck "constant'' at each point of the space-time.
Indeed, if the geometric field $\Theta(x)$ changes slowly in a cosmic scales,
then in the framework of the small domain of the laboratory we will see
an effective "constant":
$\hbar (x)\approx\hbar_\infty e^{3\Theta(x_{laboratory})}= const=\hbar$.

It can be easily seen that the Saa's model for geometric fields
$g_{\alpha\beta}$ and $\Theta(x)$ in vacuum is equivalent
to the Brans-Dicke theory \cite{BD}, \cite{Brans} in vacuum with parameter
$\omega= -{\sfrac 4 3}$.
The corresponding Brans-Dicke scalar field $\Phi = e^{-3\Theta(x)}$
in vacuum replaces the $\Theta$ field in Saa's model.
It is well known that the solutions for the scalar field in Brans-Dicke theory
outside the matter go fast to a constant \cite{BD}, \cite{Brans}.
Hence, the same property will have the $\Theta$ field in Saa's model and
the value of this field far from matter is some constant $\Theta_\infty$
which may be incorporated in a natural way into the value of Planck constant.
If we do this, we may accept the value  $\Theta_\infty\equiv 0$ as
an universal asymptotic value of the $\Theta$ field outside the matter,
and the standard experimental value of the Planck constant approximately
as an asymptotic value $\hbar_\infty$ of the new field $\hbar(x)$.

This way we reach some new interpretation of the Saa's model of gravity with
propagating torsion as a theory with variable Planck ''constant'' (VPC model)
\cite{F1}. Unfortunately, in this simplest modification of the original Saa's
model we have no SCMCP for fluids and particles, too. The usual action
principle for the action (\ref{A_total}) yields a fluid's equations of motion
\ben
(\varepsilon + p) u^\beta \nabla_\beta u_\alpha =
\left(\delta^\beta_\alpha - u_\alpha u^\beta \right)\nabla_\beta p +
{\cal F}_\alpha
\la{VPCFluEM}
\een
which are not of A-type, nor of G-type and include an additional torsion force
${\cal F}_\alpha =-2 (\varepsilon + p)
\left(\delta^\beta_\alpha - u_\alpha u^\beta \right) \partial_\beta\Theta$.

\section{The Models Based on Modified Variational Principle}

Another possibility to modify the strict Saa's model for fluids only is
to use Timan's variational principle for particles \cite{F1}.
It is not hard to see that this leads precisely to the A-type equations
of motion for relativistic fluid:
\ben
(\varepsilon + p) u^\beta \nabla_\beta u_\alpha =
\left(\delta^\beta_\alpha - u_\alpha u^\beta \right)\nabla_\beta p.
\la{AFluEM}
\een
Hence, in this modification we have SCMCP both for matter and for matter fields.
But it turns out that the Bianchi identity in this case yields the constraint
$\varepsilon^{\alpha\mu\nu\lambda} u_\mu \partial_\nu u_\lambda \equiv 0$
on the fluid motion. This constraint brings us to some interesting physical
consequences \cite{F1}  which are not studied in details at present.

There exist one more simple possibility to modify Saa's model. The combination
of the VPC model and the Timan's variational principle leads to fluid's equation
of motion with torsion force
${\cal F}_\alpha =-3 (\varepsilon + p)
\left(\delta^\beta_\alpha - u_\alpha u^\beta \right) \partial_\beta\Theta$
\cite{F1}.

\section{Spherically Symmetric Solutions in Vacuum}

Fortunately the form of the vacuum solutions for the geometric fields
$g_{\alpha\beta}$ and $\Theta$ in Saa's model
does not depend of the model of matter we use.
These solutions are simply the solutions of Brans-Dicke model with
$\omega= - {\frac4 3}$.
In Schwarzshild's coordinates the four-interval is
$ds^2 = e^{\nu}(c\,dt)^2 - e^{\lambda}(dr)^2 - r^2(d\Omega)^2$ and
the asymptotic flat, static and spherically symmetric general vacuum solutions
may be easily described as a functions of the
variable $\nu$. They depend on two additional parameters -- $\{K, a\}$:
\ben
r ={\sfrac 1 2} a e^{(3K - 1)\nu\over 2 }
\sinh ^{-1}\left(\sfrac {\rho\nu} 2\right), \nonumber \\
\nonumber \\
e^{\lambda} = \left({\sfrac {1 + \delta} 2} e^{-\rho\nu\over 2 }
+ {\sfrac {1 - \delta} 2} e^{\rho\nu\over 2}\right)^{-2},  \nonumber  \\
\nonumber \\
\Theta = {\sfrac 1 2} K \nu,
\la{VS}
\een
where $\rho = \sqrt{ 3\left(K - {\sfrac 1 2}\right)^2 + {\sfrac 1 4}}$ and
$\delta ={\sfrac {3K - 1} \rho}$.
This is the most convenient form of the vacuum solutions
 (See for details \cite{BFY}).

The same solutions in isotropic coordinates yield the four-interval in a form
\ben
ds^2 =
\left(1-{{\bf r}_{0}\over {\bf r}}\over 1+{{\bf r}_{0}\over {\bf r}}
\right)^{2\over \rho}(c\,dt)^2
- \left(1 - {{\bf r}_{0}^2 \over {\bf r}^2}\right)^2
\left(1-{{\bf r}_{0}\over {\bf r}} \over 1+
{{\bf r}_{0}\over {\bf r}}\right)^{{2\over \rho}(3K - 1)}
\left(d{\bf r}^2  + {\bf r}^2d\Omega^2\right)
\la{Is_metric}
\een
where the parameter ${\bf r}_{0}$ instead of the parameter $a$ is used.
(Note that ${\bf r}\neq r$.)

The parameter $K$ appears as an arbitrary integration constant
and presents the ratio of the magnitude of the torsion force
(as defined in \cite{F1}) and the gravitational one:
$K = { \Theta\,' / ({\sfrac 1 2} \nu\,') }$. Here and further on the prime
denotes a differentiation with respect to the Schwarzshild's variable $r$
(sometimes called "an optical radius").
In the case when $K = 0$ we have the usual torsionless Schwarzshild's solution
and $a = {\frac {{\bf r}_0} 4} \equiv r_g$ is the standard gravitational radius $r_g$.
The parameter $a$ (or ${\bf r}_0$) may take arbitrary positive values.

\section{The Spherically Symmetric Static Neutron Star
         in the Strict Saa's Model}

The main difficulty to get experimental consequences
in the models under consideration is the appearance of the new
fundamental parameter of the theory $K$ (which is constant in vacuum).
Fortunately, it turns out that its value is determined by the properties
of the usual matter. This was seen first in the model of stars \cite{BFY}
where the definite dependence of $K$ on the
total mass $M$ of the star, or on its radius $R$, as far as
on the equation of state of the star's matter was shown
via the solution of the full system of equation of the
spherically symmetric static star's state.
The parameter $a$ (or $r_0$) turns to be
related to the total mass $M$ of the star, too.

In the strict Saa's model the basic system (\ref{GFE}) for spherically
symmetric solutions in Schwarzshild's variables takes
the following normal form
\ben
{\nu\,'} = 2 \xi, \nonumber \\
{\Theta'} = S_{r}, \nonumber \\
\nonumber \\
{\xi'} = -{\sfrac \xi r} + \left ({\sfrac {2\kappa} {c^2}}\varepsilon -
{\sfrac \xi r}  - {\sfrac\kappa {c^2}}(\varepsilon  - p){r \xi} \right)
 e^{\lambda}, \nonumber \\
{S_{r}'} = -{\sfrac {S_{r}} r} + \left({\sfrac \kappa {c^2}}(\varepsilon
- p)  - {\sfrac {S_{r}} r}  -  {\sfrac \kappa {c^2}}(\varepsilon  - p)
{r S_{r}} \right) e^{\lambda},  \nonumber \\
p' = - (\varepsilon  +  p)\xi, \nonumber \\
p = p(\varepsilon),  \nonumber \\
\nonumber \\
e^{\lambda} = {{1 + 2r{\xi} - 6rS_{r}  -  3r^2{\xi}
S_{r}  +  6r^2{S_{r}}^2} \over {1 + {\sfrac \kappa {c^2}}pr^2 }}.
\la{NF}
\een
From it we obtain \cite{BFY} the following generalization
of the well known from general relativity Oppenheimer-Volkoff \cite{OppV}
system for star's equilibrium:
\ben
m_\nu' = \left(1 - \left( 1 + {\sfrac \kappa {c^2}}(\varepsilon - p)r^2
 \right)e^{\lambda} \right){\sfrac {m_\nu} r} + {\sfrac {16\pi} {c^2}}
 r^2\varepsilon
e^{\lambda}, \nonumber \\
m'_{\theta} = \left(1 - \left( 1 + {\sfrac \kappa {c^2}}(\varepsilon - p)r^2
 \right)e^{\lambda} \right){\sfrac {m_{\theta}} r}+{\sfrac {24\pi} {c^2}}r^2
(\varepsilon  - p)e^{\lambda},   \nonumber \\
p' = -{\sfrac G {c^2}}(\varepsilon + p){\sfrac {m_\nu(r)} {r^2}} \nonumber \\
p = p(\varepsilon),  \nonumber \\
 \nonumber \\
e^{\lambda} ={{1 + {2G \over c^2r} (m_\nu - m_{\theta}) - {{G^2} \over c^4 r^2}
m_{\theta}(m_\nu - {2 \over 3}m_{\theta}) } \over {1 + {\kappa \over c^2}pr^2}}
\la{OV}
\een
where $\varepsilon$ is the energy density of the star's matter, $p$ is its
pressure, and $p = p(\varepsilon)$ represents the equation of matter state.
In addition we introduce the following two {\em positive} local masses:
$m_\nu(r) = {c^2 \over 2G} r^2\nu\,'(r)$ and
$m_{\theta}(r) = 3{c^2 \over G} r^2 \Theta\,'(r)$,
$G$ being Newton gravitational constant.
For them we have $m_\nu(0) = 0$ and $m_{\theta}(0) = 0$,
because of the proper initial conditions.
After some algebra the system (\ref{OV}) supplied by these initial
conditions yields the relation
\ben
m_\nu - m_{\theta} = -  e^{A(r)}\int_0^r e^{-A(r)} (\varepsilon - 3p)dr
\la{M-M}
\een
where $A(r) = \int_0^r  {\sfrac {1 - \left(1+{\kappa \over c^2}
(\varepsilon - p)r^2 \right)e^{\lambda}} r} dr$.
The relation (\ref{M-M}) shows that ${m_\nu - m_{\theta}} \leq 0$
inside and outside the matter if ${\varepsilon - 3p} \geq 0$,
i.e. in the case of normal matter.
In other words we obtain  for $k(r)={1 \over 3} {m_{\theta}(r)\over m_\nu(r)}$
that $k\geq {1 \over 3}$ and $k$ takes a value ${1 \over 3}$ when
the matter is ultrarelativistic ($\varepsilon = 3p$).
The parameter $K$ takes its maximum value $ {1 \over 2 }$
in the case of nonrelativistic matter ($\varepsilon \gg p$).
Hence for {\em realistic} equations of state we obtain
$K\in[{1 \over 3},{1 \over 2}]$
\footnote{If we allow (following Zel'dovich \cite{Zel'dovich1}, \cite{ZN}) the
existence of some unphysical matter which breaks the usual
energy condition $\varepsilon > 3p$, i.e. if
$\varepsilon < 3p $ may take place, then in general
the vacuum value of k(r) which is just $K = k(R)$ may change its
sign passing through the zero at $\varepsilon = p$.}.

In Figure 1 we present some results of numerical calculations in the
strict Saa's model of neutron star \cite{BFY}.

Our general considerations are valid for stars with arbitrary equation of
matter state.
We chose the neutron star equation of state because in this case the nonlinear
features of the model become most transparent due to the huge matter densities,
as we know from general relativity.
The left figure correspond to the original Oppenheimer-Volkoff model
of neutron star based on the equation of matter state of non-interacting
neutron gas \cite{OppV}.
The right figure describe the same relations for the case
of more realistic Tsuruta-Cameron equation of matter state \cite{TC}.
It is seen that the M - R curves in both cases are fairly similar to these
of general relativity, but there are significant differences, too.
The maximum mass $M_{max} $ on the left figure is $ \approx 1M_{\bigodot}$,
while in general relativity the Oppenheimer-Volkoff's mass is
$M_{OV} \approx  0.7M_{\bigodot}$. The radius corresponding to the mass
$M_{\bigodot}$ is  $R \approx  4.2 km $,
while in the case of general relativity $ R_{OV} \approx  9.6 km $.
Hence, in the model under consideration the neutron
star is more compact and has a mass about $1.5$ - times greater than $M_{OV}$.
We see from the right figure that the maximum mass in the case of
Tsuruta-Cameron equation of state is
about $4.5M_{\bigodot}$ and the corresponding radius is about $7.5km$ - the
same quantities in general relativity are correspondingly
$\approx 1.6M_{\bigodot}$ and $\approx 11.5km$.
Hence, in Saa's model the interaction between the nucleons leads to an
increase in the maximum mass, as in general relativity.

Much more information about the neutron star structure and about the behavior
of all quantities in the model under consideration may be found in \cite{BFY}.
Here we will note only the main fact that relative to the predictions of general
relativity in the  strict Saa's model the torsion-dilaton field increases up
to 5 -- 6 times (depending on the equation of matter state) critical mass of
the neutron star, because it decreases the role of the gravity in the star
structure.

%---------------------------------------------------------------------------
\vskip 2truecm
\begin{figure}[htbp]
\vspace{4truecm}
\includegraphics{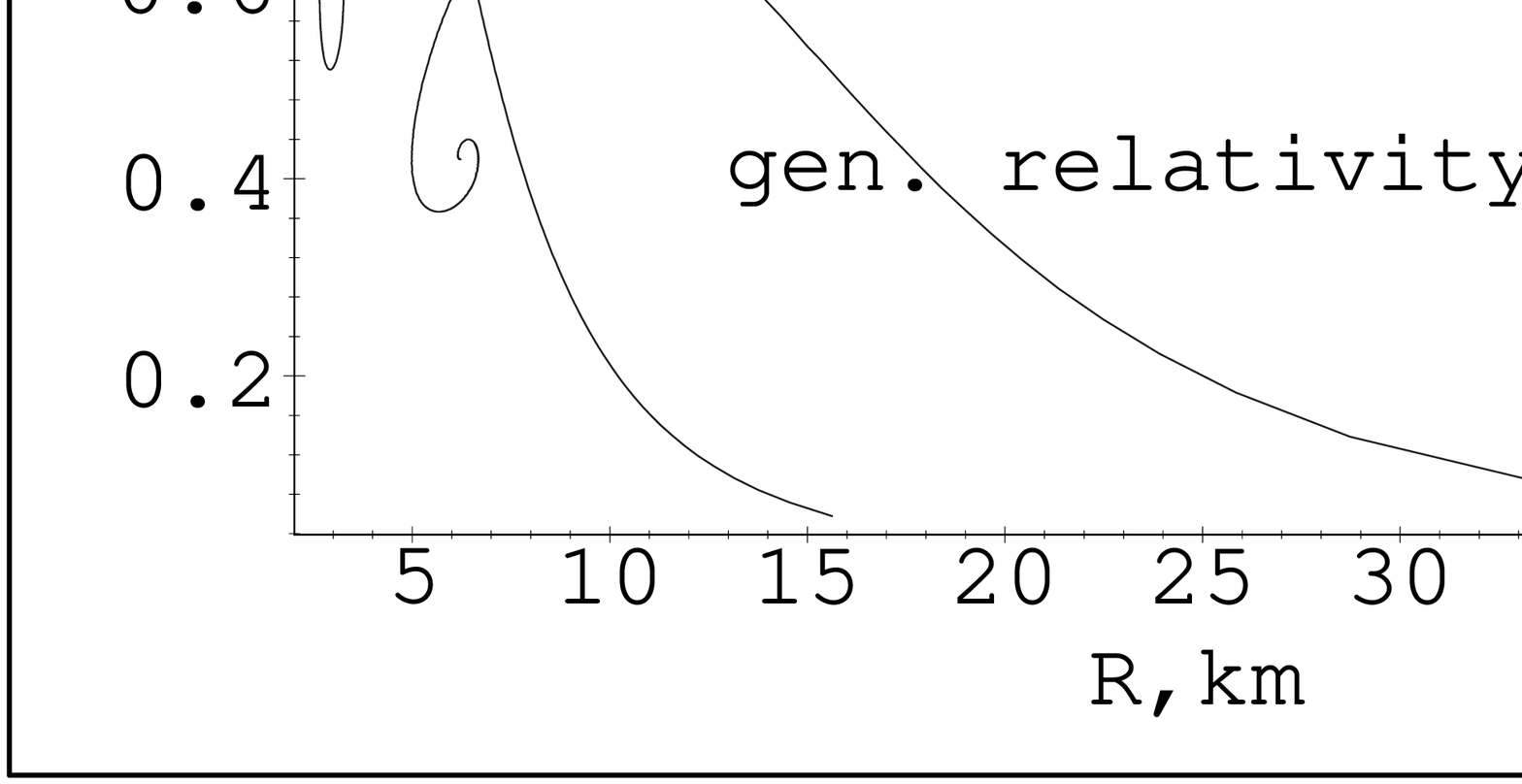}
\includegraphics{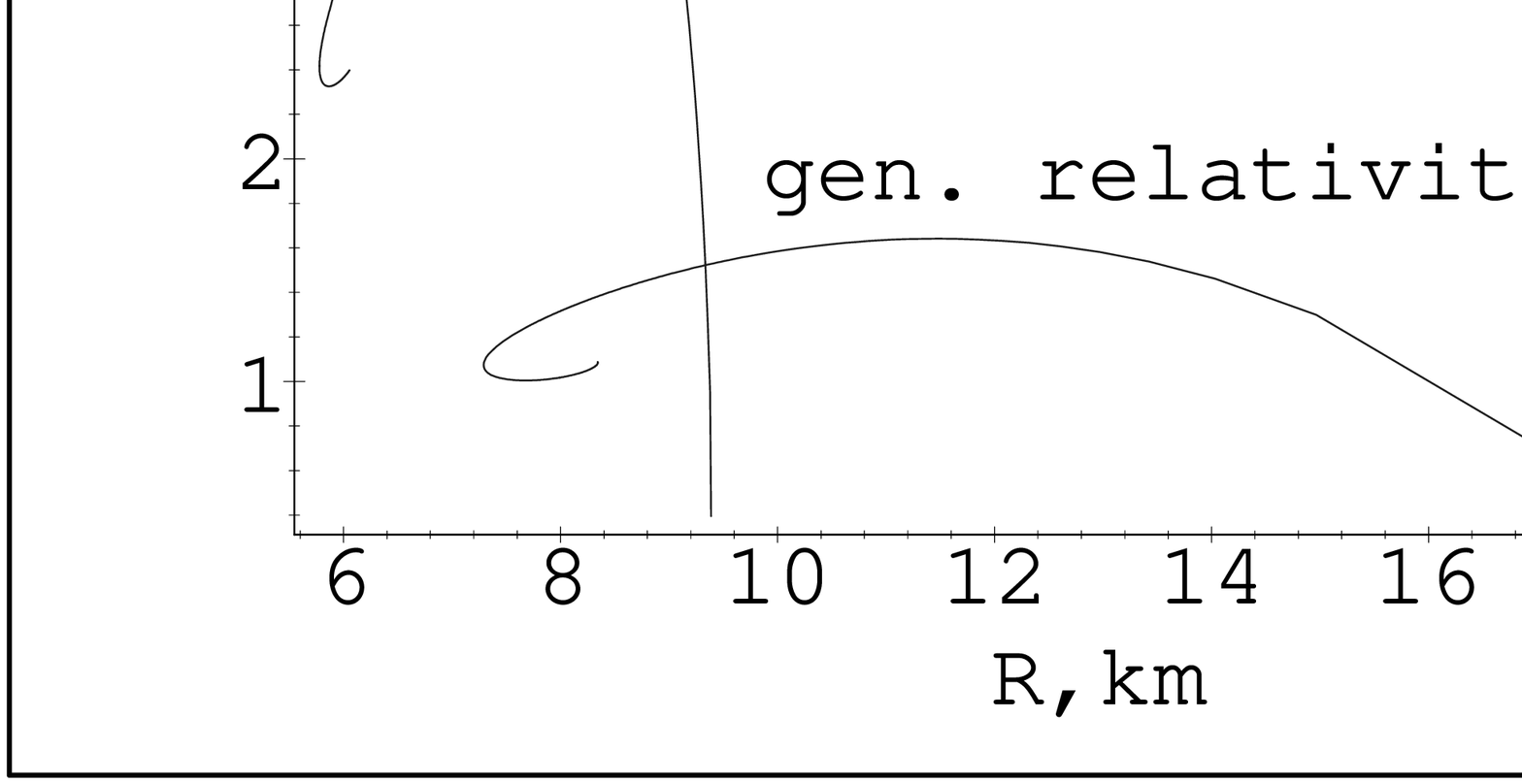}
   \vskip 0.5truecm
    \caption{Total star mass $M$ versus star radius $R$.
    Left figure represents the Oppenheimer-Volkoff model.
    Right figure -- the model with Tsuruta-Cameron equation of state.}
\vspace{.0truecm}
    \label{Fig1}
\end{figure}
%---------------------------------------------------------------------------
This observation may be instructive for all models with dilaton fields of
different kind, being under intensive investigations at present.
For example, when the interaction of the dilaton field in string theories
\cite{GSW}, \cite{Kir} with the usual matter will be known, we will be able
to consider a models of stars and to go to a real physics in string models.

\section{The Solar System Gravitational Experiments and the Models with
         Torsion-Dilaton Field}

Now we are ready to compare some of the models of gravity with propagating
torsion under consideration with the basic gravitational experiments in
solar system \cite{FY1}, \cite{FY2}. The easiest way to do this is to use
the post-Newtonian parameters \cite{Will}, i.e. to obtain the asymptotic
expansion of the spherically symmetric four-interval in a form\footnote{For
our consideration it will be convenient to use isotropic coordinates which
coincide with Schwarzshild's ones in the asymptotic region
$r \rightarrow \infty$.}:
\ben
ds^2 = \left(1 - {2M\over r} + 2\beta{M^2 \over r^2}+... \right)(c\,dt)^2
- \left(1 + 2\gamma{M\over r}+...\right)\left(dr^2  + r^2d\Omega^2\right)
\la{As}
\een
and then to look at the coefficients $\beta$ and $\gamma$ for which at present
we have tight experimental constrains \cite{Will}:
\ben
\mid\beta - 1\mid < 1*10^{-3} , \mid\gamma - 1\mid < 2*10^{-3}.
\la{Experiment}
\een

\subsection{The Strict Saa's Model}
In the strict Saa's model from formula (\ref{Is_metric}) we obtain the
asymptotic expansion \cite{FY1}:
\ben
ds^2 \approx \left(1 - {2M\over r} + 2{M^2 \over r^2} \right)(c\,dt)^2
- \left(1 + 2(1 - 3K){M\over r}\right)\left(dr^2  + r^2d\Omega^2\right).
\la{As1}
\een
This gives for the two post-Newtonian parameters: $\beta = 1, \gamma = 1 - 3K$.
Hence, the experimental restriction on the coefficient $\beta$ is fulfilled,
but the experimental restriction on the coefficient $\gamma$ leads to
the requirement $\mid K\mid < {2\over 3}*10^{-3}$
which is not consistent with the theoretical prediction
$K\in[{1 \over 3},{1 \over 2}]$ obtained for a star made from usual matter.
Hence, the strict Saa's model contradicts to the basic gravitational
experiments in the solar system.

\subsection{The VPC Model}

Because of the different dependence of the test particles lagrangian on the
torsion-dilaton field $\Theta$ \cite{F1},
for a comparison of the VPC model with the solar system gravitational
experiments it is convenient first to perform a conform transformation
$ds^2 \rightarrow d\!\stackrel{*}s\!{}^2 = e^{-6\Theta}ds^2$. Then the formula
(\ref{Is_metric}) gives the asymptotic expansion \cite{FY2}:
\ben
d\!\stackrel{*}s\!{}^2 \approx
\left(1 - {2M \over r} + {2M^2\over r}\right)(c\,dt)^2 -
\left(1 + {1\over 1- 3K} {2M\over r}\right)
\left(dr^2 + r^2d\Omega^2 \right).
\een
Hence, the two post-Newtonian parameters corresponding to the effective metric
$\stackrel{*}g_{\mu\nu}$ are
$\stackrel{*}  \beta = 1 , \, \stackrel{*}  \gamma = {1\over 1-3K}$.
Therefore to avoid contradictions with the basic experimental facts we
must have $\left| {3K \over 1 - 3K} \right| < 2*10^{-3}$.
But the consideration of a spherically symmetric stationary star in the VPC
model leads to the only possible value of the parameter $K = 1$ \cite{FY2}.
This means that in this model the torsion part of gravitational
force equals to the metric one in magnitude.
As a consequence it is impossible to fulfill the second of the experimental
restrictions (\ref{Experiment}).
Hence, the VPC model is not consistent with the basic gravitational
experiments in the solar system, too.

\subsection{The Models Based on Modified Variational Principle}

As we sow in the previous subsections of Section 8 both the strict Saa's model
and the VPC model have no SCMCP for fluids and particles
and contradict to the experimental data.
The problem with SCMCP may be solved formally making use of the Timan's
modification of the usual variational principle for particles and fluids,
as explained in Section 5. Unfortunately, the simplest models described there
are not studied in details at present. As we have mentioned, in these models
some additional restrictions on the motion of the test particles and fluids
exist. They may have an interesting physical consequences. For example,
it turns out that these models do not allow a spherically symmetric
{\em stationary} solutions for stars. At present it is not clear is this
a good, or  a bad property of these models.
In principle such unusual situation may correspond to the reality:
actually we do not see any star in a stationary state.
All stars we know, at list radiate energy in different ways and
are not in a true stationary state. It is interesting to clarify is it
possible to connect this ultimate star's radiation with
the affine geometry of the space-time.

\section{A Possible Realistic Model of Gravity with Propagating
         Torsion-Dilaton}

The results we described in the previous Sections
inspire further modification of the model with torsion-dilaton field.

\subsection{The Action for Spinless Matter and Matter Fields.}

The construction of the action for spinless matter and matter fields
is based on the following facts:

1. The real success of Saa's model with respect to the SCMCP is achieved
thanks to the insertion of the factor $e^{-3\Theta}$ into the action integral
for the matter fields (\ref{AMF}). Actually for the same purpose it is enough
to put this factor precisely into the corresponding
{\em kinetic part of the action integrals} for matter fields.

2. The consistence requirement for the semiclassical limit dictates
to put into the mass terms of the lagrangians of these fields factors of the
form $e^{-p\Theta}$ with different constants $p$ for different
massive fields \cite{F1} as was pointed out in the Section 2.

3. According to Kleinert and Pelster \cite{Kl7},
the situation with SCMCP for the relativistic spinless test particles
appears to be similar to the case of matter fields in Saa's model:
if the torsion is determined by torsion-dilaton field only,
the A-type equation of motion (\ref{AL}) may be derived via the usual action
principle from the modified action\footnote{For simplicity in this section we
will use the usual atomic units $c = \hbar = 1$ in all formulae.}:
\ben
{\cal A}_m[x]= -m\int e^{-\Theta}\sqrt{g_{\mu \nu}(x)) dx^\mu dx^\nu}=
-m\int e^{-\Theta} ds.
\la{pA3}
\een

It is not hard to obtain  that the action
\ben
{\cal A}_\mu [x]=
-\int d^4x \sqrt{|g|}\, e^{-\Theta}\left({\mu + \mu \Pi}\right)
\la{FluA}
\een
via the usual variational principle yields precisely the A-type equations of
motion (\ref{AFluEM}). For this purpose one has to take into account that
in Lagrange variables the fluid's mass density is
$\mu= \mu_0
{\frac {\sqrt{g_{\mu \nu} {\dot x}^\mu {\dot x}^\nu}}  {J(x)\sqrt{|g|}}}$,
$J(x)$ being the Jacobian of the transition from Euler to Lagrange variables
(See for details the similar calculations in \cite{F1}
and the references there in).

Our SCMCP based on the combination of 1-3 gives a definite action for massive
spinless matter fields when torsion is produced by torsion-dilaton field
only\footnote{The presence of fields with nonzero spin will yield a torsion
of more general type. In this case we need a further generalization of
the SCMCP which we reach here when only spinless matter presents.}:
\ben
{\cal A}_\phi[\phi(x)] =
\int d^4x \sqrt{|g|}\,e^{-3\Theta}\,{\sfrac 1 2}\!
\left( g^{\mu \nu} \partial_\mu \phi \partial_\nu \phi -
m^2 e^{-2\Theta}\phi^2 \right)
\la{phiA}
\een
and the A-type equation of motion:
\ben
\Box \phi + m^2 e^{-2\Theta} \phi = 0.
\la{NsfEA}
\een
The semiclassical limit of this equation is
\ben
g^{\alpha\beta} \partial_\alpha \varphi \partial_\beta \varphi
= m^2 e^{-2\Theta}
\la{NewEIC}
\een
and yields precisely to the Hamilton-Jacobi equation of
the classical particle with action (\ref{pA3}).

\subsection{The Action for Geometric Fields.}

The next problem is the construction of the action for geometric fields
$g_{\alpha\beta}$ and $\Theta$. We can use the experience we gained working
in the previous models of gravity with torsion-dilaton field after some
important remarks.

1) It is not hard to see, that the torsion-dilaton field $\Theta$
is the {\em only scalar field} which enters in the total torsion tensor
${S_{\alpha\beta} }^\gamma$ of a general affine connection,
being complete independent of the metric $g_{\alpha\beta}$.

2) If we wish to preserve general relativity as a right theory of the
metric part of the space-time geometry we have not to destroy the metric
dependence of the corresponding action. Then the only
possibility is to write down this part of the action in a form
\ben
{\cal A}_G[g_{\alpha \beta}, {S_{\alpha\beta}}^\gamma] =
-{\frac 1 {2\kappa} } \int\,F(\Theta)\, R\,\sqrt{|g|} d^4 x,
\la{AGF_F}
\een
with an arbitrary new function $F(\Theta)$.
This way we worked out at the same time the action for the torsion-dilaton
field $\Theta$, which enters in the Cartan curvature $R$ according to the
formula
$R=\stackrel{\{\}}{R} + 6\left({\stackrel{{}^{\{\}}}{\Box}} \Theta -
g^{\mu \nu} \partial_\mu \Theta \partial_\nu \Theta\right)$.
In this unique situation we do not need to put by hands additional
terms and coupling constants for the torsion-dilaton field $\Theta$,
i.e. we have a specific form of SCMCP for the very torsion dilaton field.
Hence, as in the previous models, the space-time geometry will be complete
determined by the usual properties of the matter and the interaction
of the matter with torsion-dilaton field is definitely determined by the
geometry via SCMCP, possibly supplied by some additional requirements.

3) We know from previous considerations that the choice of the new function
$F(\Theta)$ in a form $F(\Theta)= e^{-3\Theta}$ contradicts to the basic solar
system gravitational experiments. Hence, a new choice of this function
is needed. It cannot be derived via the SCMCP, because now we have to
determine in a physically acceptable way the self-interaction of the
geometrical fields $g_{\alpha\beta}$ and $\Theta$.
The SCMCP gives no instructions in this direction.

a) There exists a simple choice
$F(\Theta)= e^{-2\Theta}$ which yields a theory just in the spirit of string
theories \cite{GSW}, \cite{Kir}. It is very interesting to investigate such
model in details. In it the string dilaton appears in the torsion,
not in the Weyl's nonmetricity as it was initially proposed in \cite{SS}.
If we accept this idea then our approach has an important advantage:
the geometry and the SCMCP will imply a definite interaction of the
string-torsion-dilaton field with the usual matter. A nonminimal coupling
of the torsion-dilaton with matter fields based on some additional principles
is also possible.
We have to remind the reader that the standard string theory is not able
to make such predictions at present.

It is evident that if in this variant of theory we consider only
geometric fields and usual matter, i.e. if the total action is
${\cal A}_{total}= {\cal A}_G + {\cal A}_\mu + {\cal A}_m$
then in Einstein frame
($ g^{\mu \nu}_{Einstein} = e^{-2\Theta} g^{\mu \nu}$)
we will turn back to the usual general relativity. Hence, this variant
of theory is consistent with all known gravitational experiments concerning
usual matter \cite{Will}.
But some new effects may emerge due to the new theory
of matter fields in presence of torsion-dilaton.

b) Another very interesting possibility is to use a function
$F(\Theta)\neq e^{-2\Theta}$ which is consistent with solar system experiments.
One has to add that considering a pore dilatonic gravity with action
(\ref{AGF_F}) S. Kalyana Rama showed recently \cite{Rama}
that for a large class of functions $F(\Theta)$
we can reach cosmological models without singularities.

Maybe it will be possible to combine:\\
\noindent 1) the requirement for the existence of an universal SCMCP,
based on the
interpretation of the dilaton field as a potential of the torsion vector;\\
\noindent 2) the requirement for a right description of the solar system
experiments which give quite strong test of any theory of gravity;\\
\noindent 3) the conditions which lead to an absence of cosmological
singularities;\\
\noindent and probably \\
\noindent 4) some additional physical requirements;\\
\noindent and this way to obtain at the end some simple physically consistent
model of gravity with propagating torsion. The results of the study of these
intriguing new possibilities will be described in next papers.

Another obvious and necessary step will be the investigation of the physical
relevance of other field degrees of freedom which enter in the Cartan torsion
tensor and may be related with non-zero-spin matter.
The corresponding considerations will be represented somewhere else.

\bigskip
\bigskip

\noindent{\Large\bf Acknowledgments}
\bigskip

This work has been partially supported by
the Sofia University Foundation for Scientific Researches, Contract~No.~245/98,
and by
the Bulgarian National Foundation for Scientific Researches, Contract~F610/98.

The author is grateful to the leadership of the
Bogoliubov Laboratory of Theoretical Physics, JINR, Dubna, Russia
for hospitality and working conditions during his stay there in
the summer of 1998, as far as to the organizers of the
XI International Conference Problems in Quantum Field Theory,
Dubna, Russia,  July 13-17, 1998 for the possibility to join this Conference
and to give this talk.

The author also wishes to express his thanks for the stimulating discussions
of different parts of the present talk to S.~Yazadjiev, T.~Boyadjiev,
B.~M.~Barbashov, V.~V.~Nesterenko, A.~A.~Zheltukhin,
A.~Pelster, V.~de Alfaro and M.~Cavaglia.

\end{document}